# Assembly line balancing considering stochastic task times and production defects

[1]Gazi Nazia Nur, [1]*Mohammad Ahnaf Sadat and [2]Basit Mahmud Shahriar

[1] Iowa State University, Ames, Iowa, United States of America
[2] IPE, Military Institute of Science and Technology, Mirpur Cantonment, Dhaka-1216, Bangladesh.

*E-mail of the corresponding author: sadat@iastate.edu

**ABSTRACT**

In this paper, we address the inherent limitations in traditional assembly line balancing, specifically the assumptions that task times are constant and no defective outputs occur. These assumptions often do not hold in practical scenarios, leading to inefficiencies. To address these challenges, we introduce a framework utilizing an "adjusted processing time" approach based on the distributional information of both processing times and defect occurrences. We validate our framework through the analysis of two case studies from existing literature, demonstrating its robustness and adaptability. Our framework is characterized by its simplicity, both in understanding and implementation, marking a substantial advancement in the field. It presents a viable and efficient solution for industries seeking to enhance operational efficiency through improved resource allocation.

## 1. INTRODUCTION

The core function of a production system is to effectively coordinate tasks and resources to convert raw materials into finished products. These systems are broadly classified into two categories based on the flow of production: intermittent systems and continuous systems [1]. With the push for industrialization and the pursuit of economies of scale, industries are opting for continuous manufacturing. In such systems, products move along a line from one workstation to the next, with a specified number of workers completing their assigned tasks at each workstation. This arrangement is commonly referred to as an assembly line, where each workstation has a set number of tasks, each with its own time requirement and order of execution [2]. Despite the diversity of tasks at each workstation, all workstations are given the same maximum duration to complete all their respective tasks. Throughout the paper, we refer to this duration as cycle time (please see paced assembly lines in Becker and Scholl [3] for more details about cycle time).

In the context of assembly line balancing, the precise assignment of tasks to workstations is crucial to avoid idle time, which occurs when workers are not actively engaged but are waiting for others to finish their tasks within the cycle time before a new cycle can begin. Tasks are assigned to various workstations through methods of parallelization and serialization, aiming to build an assembly line that minimizes idle time and labor costs [4]. Operations researchers strive to optimize these assembly lines by reducing the number of workstations, evenly distributing the workload, decreasing cycle time, and enhancing overall efficiency. These optimization problems often involve treating specific aspects as variables while keeping others as fixed parameters, leading to the creation of a diverse family of optimization problems. Additionally, within this family of problems, distinctions can be made based on the nature of task times, which can be either deterministic or stochastic [4]. In deterministic scenarios, task times are assumed to remain constant, whereas in stochastic scenarios, task times are considered to follow a distribution.

A substantial body of research has delved into assembly line balancing, exploring various areas such as the development of new mathematical models, the derivation of initial solutions using heuristic techniques, and the refinement of heuristic methods. For a comprehensive understanding of the literature on assembly line balancing, readers are encouraged to refer to the literature reviews [5], [6].

Despite this extensive research, to our best knowledge, very few studies [7] have addressed another critical aspect of manufacturing — "defects" in assembly line balancing. Defects are inherent in manufacturing systems, particularly in scenarios involving manual labor, influenced by various human factors such as motivation, work environment, and mental and physical stress [8]. The oversight of defect generation during assembly line balancing can compromise efficiency.

Moreover, failure to address defects promptly upon their generation can lead to unnecessary work on defective items. Some defects may be irreparable, while rectifying others at a later stage may require significant additional effort. As noted by Boysen et al. [6], defects should be managed in line, not after completing all tasks. Additionally, Sadat et al. [9] demonstrated that in-process quality control has the capability to enhance productivity.

Taking these factors into account, this paper aims to develop a simple framework for assembly line balancing that integrates defect generation throughout the production process. Furthermore, within this framework, we advocate that defect management should be conducted within the same workstation where defects arise, rather than being handled separately. Through this approach, we aim to align our model more closely with real-world scenarios and achieve greater efficiency compared to existing literature.

## 2. PROBLEM DESCRIPTION

In the context of human-operated work environments, existing assembly line balancing algorithms found in the literature are inadequate due to their oversight of inherent uncertainties related to production quality, particularly in defect generation. This neglect has the potential to compromise the efficiency of the assembly line. The aim of this study is to develop a simple framework for balancing assembly lines that account for the occurrence of defects during production. Additionally, in order to better reflect real-world conditions, we consider that task's processing times are stochastic.

## 3. PROPOSED FRAMEWORK

The framework for assembly line balancing encompassing stochastic processing times and defect generation is based on the following set of assumptions:
  i. The processing time of the tasks follows normal distributions.
     -This assumption is based on the idea that the processing times for tasks on the assembly line vary around a mean (average) time with a certain standard deviation.
  ii. Defects can occur during any production task and can be modeled using a Poisson distribution.
     - This assumption is based on the idea that defects arise randomly and independently across different tasks with a constant average rate. The Poisson distribution is ideal for modeling the number of defects because it describes the probability of a certain event occurrence number within a fixed time interval. This framework rests on the assumption that these events occur with a known constant mean rate and are independent of the previous occurrences. This is particularly applicable in manufacturing environments where the defect rate is consistent over time, and each defect occurrence is independent of the others.
  iii. After each task, the work-in-progress items may require defect inspection. We have factored this inspection time into the task's processing time.
     -This assumption reflects our hypothesis that identifying defects is crucial for improving line-balancing efficiency. It also acknowledges the importance of quality checks within the production process to ensure that each stage meets specified standards before advancing to the next.
  iv. Any necessary rework is performed at the same workstation and by the same worker where the defect originated.
     - This simplifying assumption is based on the premise that having the same worker address any reworks minimizes the learning curve and communication errors since the worker is already familiar with the specific item and the nature of the defect. Additionally, this approach prevents the need to move defective items across different stations, thereby reducing cycle times and enhancing the overall efficiency of the line-balancing.

This proposed framework is structured into two segments. In the first segment, we determine the distribution parameters related to the "adjusted processing time", a concept that will be thoroughly discussed later in this section. And in the second segment, we apply any line-balancing algorithm using these adjusted processing times (such as Moodie and Young [10], Integer Linear Programming (ILP), and others) to balance the production line and then evaluate the efficiency of the production line using the proposed efficiency measure. The "adjusted processing time" metric integrates the processing time of tasks and the revised reworking time. The term "revised" is used deliberately to emphasize the inclusion of the probability of defect generation with the original reworking time (see equation (1)). The original reworking time itself comprises two elements: dismantling time and processing time of the task. Therefore, the adjusted processing time for any task can be expressed as equation (2):

Revised reworking time = Probability of defects generated in the task × Original reworking time
 = Probability of defects generated in the task × (Dismantling time + Processing time of the task) (1)

Adjusted processing time = Processing time of the task + revised reworking time

$$= (\mu + Z_{p_1}\sigma) + \left(\frac{K_{p_2}}{L} \times ((\mu_d + \mu) + Z_{p_1}\sqrt{\sigma_d^2 + \sigma^2})\right)$$ (2)

Here,
$\mu$ = mean processing time of the task
$\sigma$ = standard deviation of the task's processing time
$Z_{p_1}$ = the Z-value corresponding to the $p_1$-th percentile
$L$ = lot size
$K_{p_2}$ = the number of defects generated for the $p_2$-th percentile value = $F^{-1}\left(\sum_{i=0}^{k} \frac{(vL)^i}{i!} e^{-vL}\right)$
$v$ = the probability of a unit being defected

The proposed framework consisting of two segments is presented below:

Segment 1: Determining the distributional parameters

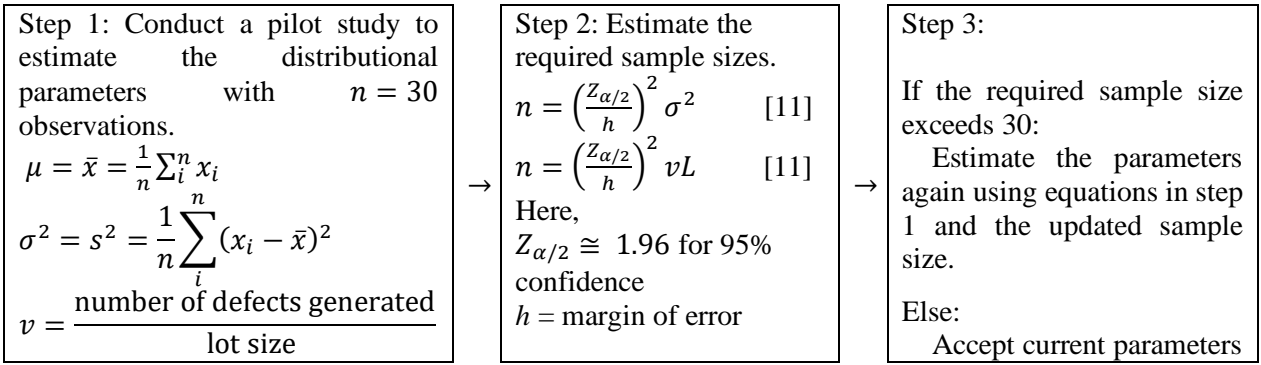

Segment 2: Applying line balancing algorithm and evaluating efficiency

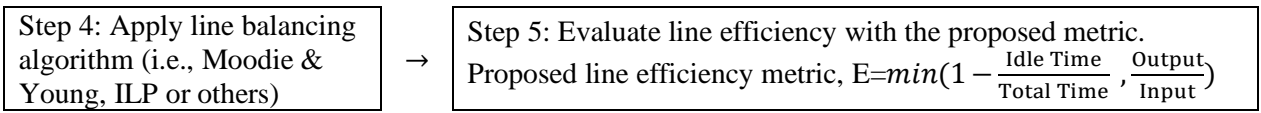

The first component of the proposed line efficiency metric $\left(1 - \frac{\text{Idle Time}}{\text{Total Time}}\right)$ is a popular measure in deterministic line balancing scenarios [12]. The second component $\left(\frac{\text{Output}}{\text{Input}}\right)$ extends efficiency evaluation to diverse situations. This measurement can also indicate the effectiveness of line balancing. By adopting the proposed efficiency metric, we suggest that production line efficiency should be determined by selecting the minimum value, whether it is due to lost productive time or finished product defects.

## 4. METHODOLOGY

Our lack of access to production environments prevents us from collecting real-life assembly line balancing data. Therefore, we obtained two deterministic assembly line balancing problems from existing literature sources. One is a more theoretical problem outlined by Hoffman [13], consisting of 9 tasks, while the other is a practical scenario involving a shirt sewing process with 15 tasks, as described by Kayar and Akyalçin [14]. These problems exclude random variables such as processing time variability and defect generation, and replicating these scenarios is beyond the scope of our paper.

As a result, we were unable to execute segment 1 (steps 1, 2, and 3) of the proposed framework. Instead, we made assumptions regarding the distributional parameters that would typically result from step 3 of segment 1. Utilizing these assumed parameters and selecting the 50th percentile for both normal and Poisson distributions, we determined the adjusted processing time.

Subsequently, we applied Moodie and Young [14] as well as ILP (optimization model for minimizing workstation for a given cycle time, SALBP-1 from Boysen et al. [6]) to balance the assembly lines. Both methods are explained briefly in Appendix A and B, respectively. Following line balancing, we conducted simulations (implemented in Python) of the balanced lines 100 runs to evaluate efficiency based on our proposed metric. The results obtained from these simulations are reported in our study. Additionally, we varied the percentiles for the normal and Poisson distributions to generate a surface plot for further analysis.

## 5. NUMERICAL EXAMPLES

We sourced two deterministic assembly line balancing problems from the literature. The first one, the 9-task problem (outlined in **fig. 1.**) along with its associated mean processing time (detailed in row 2 of **Table 1**), was directly obtained from Hoffman [3]. However, we made assumptions for the remaining data points in Table 1—such as the standard deviation of processing time, mean and standard deviation of dismantling time, and the number of defects generated within a lot size of 50. Using Equation 2, with both the 50th percentile of the normal distribution and the Poisson distribution, we calculated the adjusted processing time, which is also listed in Table 1.

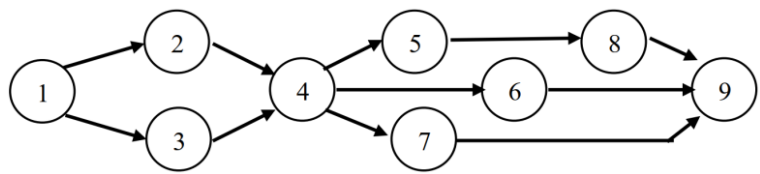

**Figure 1**: 9-task problem (Hoffman [13])

**Table 1**: Adjusted processing time and other associated data for the 9-task problem

| Tasks | | 1 | 2 | 3 | 4 | 5 | 6 | 7 | 8 | 9 |
|---|---|---|---|---|---|---|---|---|---|---|
| Processing time (min) | Mean [13] | .5 | .3 | .4 | .5 | .4 | .5 | .1 | .4 | .6 |
| | Std Dev | .1 | .1 | .1 | .1 | .1 | .1 | .02 | .2 | .1 |
| Dismantling time (min) | Mean | 1.2 | 1.4 | 1.5 | 1.6 | 1.5 | 1.6 | 1 | 1.5 | 1.6 |
| | Std Dev | .2 | .1 | .2 | .2 | .2 | .2 | .1 | .2 | .25 |
| Mean defect generated (units) | | 10 | 12 | 14 | 11 | 12 | 8 | 9 | 13 | 8 |
| Adjusted processing time (min) | | .84 | .71 | .93 | .96 | .86 | .84 | .00 | .89 | .95 |

Similarly, we acquired another assembly line balancing problem, namely the shirt sewing problem involving 15 tasks, from Kayar and Akyalçin [4]. This problem is illustrated in **fig. 2**. and detailed in **Table 2**. For this particular problem, we utilized a lot size of 100 items. Once again, we employed the 50th percentile for both the normal and Poisson distributions to compute the adjusted processing time.

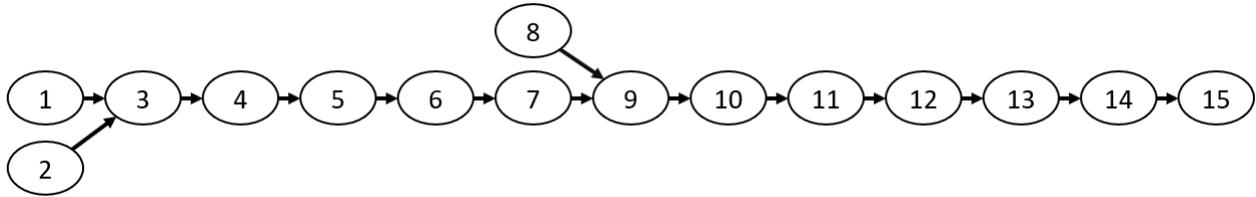

**Figure 2**: Shirt sewing problem from Kayar and Akyalçin [14]

Similar to the previous problem, we have obtained the shirt sewing problem with 15 tasks from Kayar and Akyalçin [14] presented in **fig. 2**. and **table 2**.

**Table 2**: Adjusted processing time and other associated data for the shirt sewing problem

| Tasks | | 1 | 2 | 3 | 4 | 5 | 6 | 7 | 8 | 9 | 10 | 11 | 12 | 13 | 14 | 15 |
|---|---|---|---|---|---|---|---|---|---|---|---|---|---|---|---|---|
| Processing time (min) | Mean [14] | .29 | .08 | .34 | .25 | .35 | .08 | .36 | .09 | .43 | .42 | .13 | .33 | .06 | .25 | .45 |
| | Std Dev | .10 | .02 | .12 | .11 | .14 | .02 | .12 | .02 | .13 | .15 | .04 | .11 | .01 | .09 | .17 |
| Dismantling time (min) | Mean | .38 | .15 | .30 | .35 | .45 | .12 | .26 | .12 | .33 | .29 | .15 | .24 | .13 | .21 | .30 |
| | Std Dev | .11 | .07 | .02 | .12 | .12 | .04 | .11 | .03 | .12 | .11 | .06 | .10 | .05 | .09 | .10 |
| Mean defect generated (units) | | 10 | 12 | 14 | 11 | 12 | 08 | 09 | 13 | 08 | 12 | 14 | 09 | 12 | 13 | 13 |
| Adjusted processing time (min) | | .36 | .11 | .43 | .32 | .45 | .1 | .42 | .12 | .49 | .51 | .17 | .38 | .08 | .31 | .55 |

## 6. RESULTS

At the end of this section, we will present the experimental results for both problems in tabular format. For clarity and conciseness, we will only walk through the results of the shirt sewing problem [14] step by step.

Initially, we applied the Moodie and Young method [10], as well as the ILP [6] separately, assuming a cycle time of 2 minutes per unit based on the data provided in **Table 2**. The resulting balanced assembly line is depicted in **Table 3.**

**Table 3**: Distribution of tasks to workstations for the shirt sewing problem [14]

| | Workstation 1 | Workstation 2 | Workstation 3 | Workstation 4 | Workstation 5 |
|---|---|---|---|---|---|
| Moodie and Young | 1→8→2→3 | 4→5→6 | 7→9 | 10→11→12→13 | 14→15 |
| ILP | 1→2→3 | 4→5 | 6→7→8→9 | 10→11→12 | 13→14→15 |

After that, we assumed the production lot size as 100, indicating that at any given time, there would be 100 work-in-progress items present at each workstation. In configuring our balanced assembly line, we allocated a cycle time of 2 minutes per shirt, resulting in a total cycle time of 200 minutes for each lot. To evaluate the efficiency of the balanced assembly lines, we carried out simulations for 100 lots (100 runs), using stochastic data from **Table 2**. Utilizing our defined efficiency metric, we observed that both assembly lines—optimized using the Moodie and Young method and the ILP algorithms—attained an efficiency score of 0.80.

By adjusting the percentiles within both the normal and Poisson distributions, we can achieve various assembly line balances using the Moodie and Young algorithm. Evaluating the efficiency of these balances allows us to generate a surface plot, as shown in **fig. 3**. Interestingly, peak efficiency is not found at a single point but occurs across a range of combinations of processing time and defect generation percentiles. Furthermore, the figure reveals that assigning higher percentile values to defects and processing times does not necessarily improve the assembly line's efficiency.

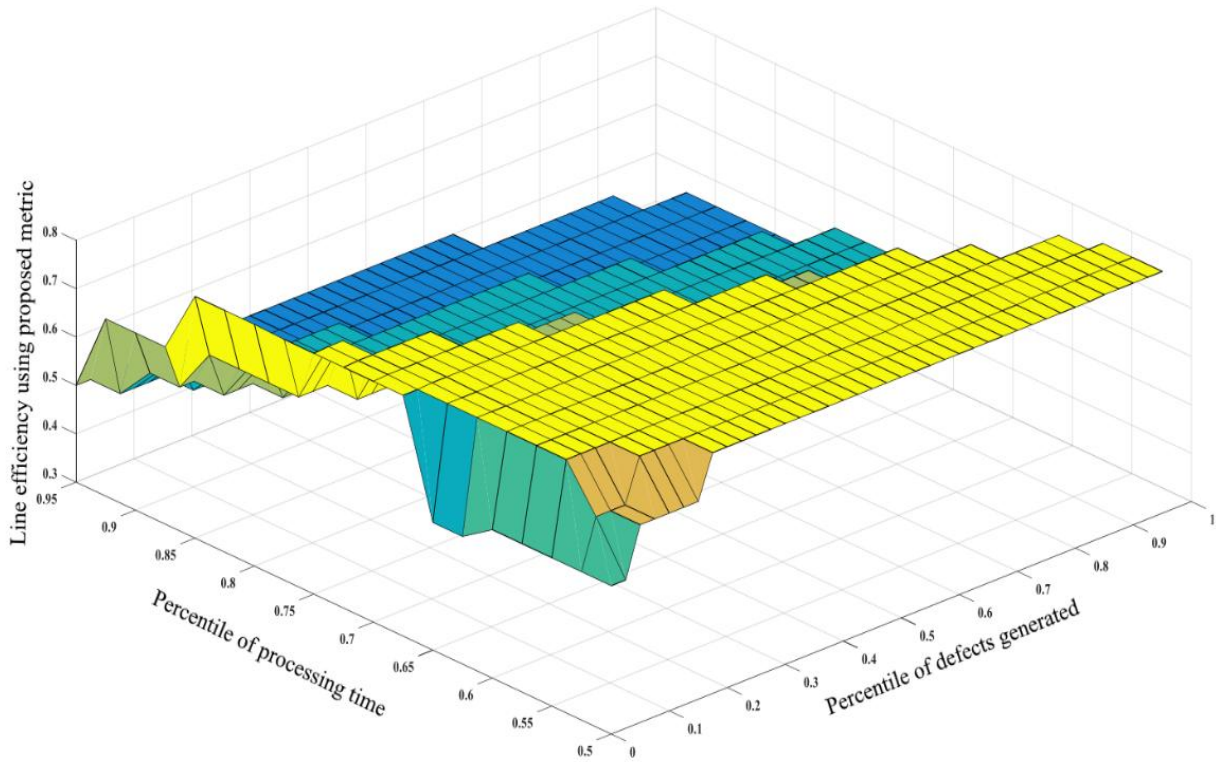

**Figure 3:** Surface plot of the shirt sewing problem from Kayar [14] using Young & Moodie method

Adopting the approach outlined for the shirt sewing scenario, we assessed the efficiencies of both the 9-task problem and the shirt sewing problem across various cycle times. For performance comparison with the deterministic model, we included the 0th percentile for defect generation, thus disregarding the potential for defects during the line balancing process. For all evaluations, we utilized the 50th percentile of processing time to calculate the adjusted process time. The results of these analyses are tabulated in **Table 4.**

**Table 4:** Impact of incorporating defect generation on assembly line balancing performance

|  |  | Moodie & Young [10] | | | | Integer Linear Programming [6] | | | |
|  |  | $0^{th}$ percentile of defect generation | | $50^{th}$ percentile of defect generation | | $0^{th}$ percentile of defect generation | | $50^{th}$ percentile of defect generation | |
| Problem | Cycle time (min) | Required workstations | Proposed efficiency | Required workstations | Proposed efficiency | Required workstations | Proposed efficiency | Required workstations | Proposed efficiency |
|---|---|---|---|---|---|---|---|---|---|
| 9-task [13] | 100 | 2 | 0.26 | 4 | 0.69 | 2 | 0.23 | 4 | 0.72 |
|  | 75 | 3 | 0.40 | 4 | 0.90 | 3 | 0.42 | 4 | 0.66 |
|  | 50 | 4 | 0.57 | 8 | 0.69 | 4 | 0.4 | 8 | 0.61 |
| Shirt sewing [14] | 200 | 3 | 0.28 | 3 | 0.80 | 3 | 0.8 | 3 | 0.8 |
|  | 150 | 3 | 0.34 | 4 | 0.79 | 3 | 0.66 | 4 | 0.79 |
|  | 100 | 5 | 0.75 | 6 | 0.79 | 5 | 0.29 | 6 | 0.76 |

**Table 4** shows that implementing the framework discussed in this paper generally leads to improved efficiency across nearly all cases. However, it also requires an increase in the number of workstations. It is important to note that while the efficiency achieved with the proposed method surpasses that of traditional approaches, it does not reach 100%. Often, efficiencies are below 80%, primarily due to the stochastic nature of the tasks involved. To address this, dynamic task allocation might be considered, but such an approach can introduce considerable complexity in manufacturing management. Therefore, we have decided not to include this option in the current study and instead suggest exploring it in future research.

## 7. CONCLUSION

In this study, we introduced a modification to processing time, termed "adjusted processing time," which combines the impact of defects produced and stochastic task time. The adjusted processing time can be balanced using any deterministic methodology. The strength of this framework lies in its simplicity and ease of implementation. We also proposed a new metric for measuring the efficiency of a balanced assembly line, advocating for efficiency to be assessed based on the minimum value derived from either lost production time or the incidence of defects in finished products. Simulation results suggest that the proposed framework excels in stochastic environments, demonstrating significant efficiency improvements across all cases examined. Industries stand to benefit from implementing this framework in line balancing, especially in scenarios where there is an opportunity to add more workstations.

Regarding limitations, we recognize that our study did not account for the impact of worker fatigue and learning curves. Additionally, we acknowledge that the efficiency of the assembly line may not reach ideal levels, particularly in scenarios where there is significant variability in task times and defect generation. Dynamic task allocation could potentially address this issue, but it may also introduce complexities in managing assembly line balancing. Therefore, a careful evaluation of the trade-off between complexity and benefits is necessary. Future research could focus on addressing these limitations by incorporating considerations of worker fatigue and learning curves, as well as exploring strategies for dynamic task allocation.

**Appendices**

Appendix A: Moddie and Young [10]

In this appendix, we will present the heuristic method offered by Moddie and Young [10]. This heuristic method requires a precedence matrix to be constructed from the task precedence relationship. For generating the task precedence matrix, we followed [13], and the following is a direct excerpt of [13]:

> "This is a square matrix, consisting of zeros and ones, in which the rows are labeled with consecutive element number and the column are labeled in the same order. Entries in the matrix are as follows:
> 1. If the element of row i immediately precedes the element of column j, a 1 is placed in row i, column j.
> 2. All other entries are zero"

The precedence matrix for the 9-task problem [13] in Figure 1 is below:

| Column\Row | 1 | 2 | 3 | 4 | 5 | 6 | 7 | 8 | 9 |
|---|---|---|---|---|---|---|---|---|---|
| 1 |   | 1 | 1 |   |   |   |   |   |   |
| 2 |   |   |   | 1 |   |   |   |   |   |
| 3 |   |   |   | 1 |   |   |   |   |   |
| 4 |   |   |   |   | 1 | 1 | 1 |   |   |
| 5 |   |   |   |   |   |   |   | 1 |   |
| 6 |   |   |   |   |   |   |   |   | 1 |
| 7 |   |   |   |   |   |   |   |   | 1 |
| 8 |   |   |   |   |   |   |   |   | 1 |
| 9 |   |   |   |   |   |   |   |   |   |

After developing the precedence matrix, we followed the following Moodie & Young [10] algorithm:

Step 1: Start with $i = 1$
Step 2: Generate the precedence matrix, and create a list of schedulable operations (the tasks whose columns are all 0)
Step 3: Select the task with longest processing time from the schedulable operation list and assigned It to the works station $i$
Step 4: Calculate the residual time of the workstation $i$ as 'cycle time – processing time of assigned operation'
Step 5: Delete the row of the assigned operation in the precedence matrix and generate a new list of schedulable jobs. If no schedulable jobs available, then STOP.
Step 6: Check if the residual cycle time > any of the schedulable operations processing time
  If true, return to step 3 and again continue
  If false, increase i by 1, reset cycle time, return to step 3, and again continue

Appendix B: ILP for SALBP-1

In this appendix, we first clarify what SALBP-1 is, and then we clarify the ILP used in this paper.

Please refer to the introduction of this paper, where we discussed how treating certain aspects of line balancing as variables and others as parameters creates a family of line-balancing problems. This family is referred to as SALBP, which stands for Simple Assembly Line Balancing. The first variant of this family, SALBP-1, focuses on reducing the number of workstations for a given cycle time.

The ILP used for SALBP-1 is a sight modification of [6] and is as follows:

Parameters:

$C_t$ = Cycle time
$I$ = The set of tasks
$J$ = The set of workstations
$m$ = The maximum number of workstations

Variables:

$$x_{ij} = \begin{cases} 1, \text{if task } i \text{ is assigned to workstation } j \\ 0, \text{otherwise} \end{cases}$$

$$y_i = \begin{cases} 1, \text{if workstation } i \text{ is used} \\ 0, \text{otherwise} \end{cases}$$

Model:

$$\text{Min,} \sum_{j=1}^{m} 2^j \times y_j$$

Subject to,

$$\sum_{j=1}^{m} x_{ij} = 1 \quad \forall i \in I$$

$$\sum_{i=1}^{n} t_i \times x_{ij} \leq C_t \times y_j \quad \forall j \in J$$

$$\sum_{j=1}^{J} j \times x_{ij} \leq \sum_{j=1}^{j} j \times x_{kj} \quad \forall\, i < k$$

$x_{ij}$ and $y_j$ are binary $\quad \forall i \in I, \forall j \in J$

The underlying idea of this integer linear mathematical model is to make assignments in new workstations extremely costly. The first set of constraints ensures a single operation must be performed once. While the second set of constraints ensures no overload occurs in a station. The third set of constraints maintains the precedence constraints. The maximum number of workstations, *m* needs to be calculated before or could be simply taken as the total number of jobs.